\documentstyle[12pt]{article}

\makeatletter
\def\la{\mathrel{\mathpalette\gl@align<}}
\def\ga{\mathrel{\mathpalette\gl@align>}}
\def\gl@align#1#2{\lower.6ex\vbox{\baselineskip\z@skip\lineskip\z@\ialign{$\m@th
#1\hfil##\hfil$\crcr#2\crcr\sim\crcr}}}
\makeatother

\def\thebibliography#1{\section*{REFERENCES\markboth
 {REFERENCES}{REFERENCES}}\list
 {}{\settowidth\labelwidth{[#1]}\leftmargin\labelwidth
 \advance\leftmargin\labelsep
 \usecounter{enumi}}
 \def\newblock{\hskip .11em plus .33em minus .07em}
 \sloppy
 \sfcode`\.=1000\relax}

\title{The influence of Galactic wind upon the star formation histories
of Local Group galaxies}
\author{H. Hirashita\thanks{E-mail: hirasita@kusastro.kyoto-u.ac.jp},
H. Kamaya, S. Mineshige \\
{\it Department of Astronomy, Kyoto University,} \\
{\it Sakyo-ku, Kyoto 606-01,
Japan}}
\date{Accepted by {\sl MNRAS}}

\begin{document}

\maketitle

\begin{abstract}
We examine the possibility that ram pressure exerted by
the galactic wind from the Galaxy could have stripped
gas from the Local Group dwarf galaxies, thereby
affecting their star formation histories.
Whether gas stripping occurs or not depends on
the relative magnitudes of two counteracting forces
acting on gas in a dwarf galaxy:
ram pressure force by the wind and
the gravitational binding force by the
dwarf galaxy itself.
We suggest that the galactic wind could have stripped
gas in a dwarf galaxy located within the distance of
$R_{\rm c}\simeq 120(r_{\rm s}/1\,\mbox{kpc})^{3/2}
({\cal E}_{\rm b}/10^{50}\,\mbox{erg}
)^{-1/2}$ kpc (where $r_{\rm s}$ is the
surface radius and
${\cal E}_{\rm b}$ is the total binding energy of the
dwarf galaxy, respectively) from the
Galaxy within a timescale of Gyr, thereby preventing
star formation
there. Our result based on this Galactic wind model
explains the recent observation that
dwarfs located close to the Galaxy
experienced star formation only in the early phase of
their lifetimes, whereas distant dwarfs are still undergoing
star formation. The present star formation in the Large
Magellanic Cloud can also be explained through our
Galactic wind model.
\end{abstract}

\section{INTRODUCTION}

It is widely accepted that the hot interstellar gas, mainly
originating from supernova explosions, will eventually
escape from the galactic disk as galactic wind
(Shapiro \& Field 1976;
Habe \& Ikeuchi 1980; Tenorio-Tagle \& Bodenheimer 1988;
Norman \& Ikeuchi 1989). Such a wind
can have various effects on the disk-halo system.
For example, it can supply energy and gas to the halo
(Chevalier \& Oegerle 1979; Habe \& Ikeuchi 1980;
Li \& Ikeuchi 1992).

We, here, consider if
ram pressure exerted by the galactic
wind from the Galaxy could have stripped gas from the
Local Group dwarf galaxies. Such ram pressure stripping may
influence star formation histories of the dwarfs.
The effects of ram pressure are extensively discussed by
Portnoy, Pistinner, \& Shaviv (1993).

We summarize various properties of Local Group dwarf galaxies
in Table 1, where $R$ is the distance from the Galaxy
(the galactocentric distance), and
$r_{\rm s}$ and ${\cal E}_{\rm b}$
are the surface radius and the binding energy calculated from
the observations of velocity dispersions, respectively. In
the fifth column of
SF (star formation), we distinguish the following two
categories:

\noindent
A: Dwarf galaxies in which the stellar
population is dominated by
the initial bursts of star formation;

\noindent
B: Dwarf galaxies which experienced
recent star formation, as well.

Van den Bergh (1994) has asserted that the star formation
histories of Local Group
dwarf galaxies correlate with the distance
from the Galaxy ($R$): Dwarf galaxies near
the Galaxy, such as Ursa Minor and Draco, only experienced
star formation in the early phase of their
lifetimes ($\sim 12$ Gyr), while there is observational
evidence for more
recent (or present) star formation
in distant dwarfs (see van den Bergh 1994
and references therein). He also implied that star formation
in dwarfs suffers great difficulty
in the existence of mass flow
from the Galaxy.

\begin{table}
{\small
  \caption{Local Group dwarf galaxies. See text for the definitions.}
  \begin{tabular}{lccccccc}
Name & $R$(kpc) & $r_{\rm s}$(kpc) &
${\cal E}_{\rm b}(10^{50}\mbox{erg})$ & SF & $R_{\rm c}$(kpc)
& $t_{\rm cross}$(Gyr) & Refernces
\footnote{(1) van den Bergh 1994; (2) Saito 1979a; (3) Da Costa 1984;
(4) Mighell \& Rich 1996; (5) Carignan, Demers, \& C\^{o}t\'{e} 1991;
(6) van de Rydt, Demers, \& Kunkel 1991;
(7) Lo, Sargent, \& Young 1993; (8) Fisher \& Tully 1979.} \\
Ursa Minor & 65 & 1.38 & 0.121 & A & 550 & 0.2 & 1, 2 \\
Draco    & 76 & 0.795 & 0.136 & A & 240  & 0.2 & 1, 2 \\
Sculptor & 78 & 1.70  & 20.9  & B & 50 & 0.2 & 1, 2, 3 \\
Sextans  & 82 & $\cdots$ & $\cdots$ & A & $\cdots$ & 0.2 & 1 \\
Fornax   & 133 & 3.89 & 86.6  & B & 100   & 0.4 & 1, 2 \\
Leo II   & 217 & 0.717 & 0.665 & B & 90  & 0.6 & 1, 2, 4 \\
Leo I    & 277 & 1.20 & 7.19  & B & 50   & 0.8 & 1, 2   \\
Phoenix  & 390 & 1 & 0.8 & B & 140        & 1 & 1, 5, 6 \\
DDO 210  & 794 & 0.7 & 10 & B & 20       & 2 & 1, 7, 8
\end{tabular}
}
\end{table}

The plan of this Letter is as follows.
First of all, in the next section, we define the critical
radius $(R_{\rm c})$ in such a
way that a dwarf galaxy within $R_{\rm c}$ from the
Galaxy suffers ram pressure stripping of the wind, and
derive the general expression for $R_{\rm c}$. We
then discuss in \S 3 the interpretation of the
observations by
van den Bergh (1994) and one exceptional case, the Large
Magellanic Cloud, whose star formation histories are also
explained with our galactic wind model. In the final section,
we shall
conclude that our hypothesis can qualitatively account
for the observation.

\section{THE CRITICAL RADIUS FOR GAS STRIPPING}

In this section we compare ram pressure
force by the wind and gravitational binding force
by a dwarf galaxy itself. We shall then evaluate the
critical radius, $R_{\rm c}$, within which the ram
pressure exceeds the gravitational force.

First we evaluate the mass loss rate of the Galaxy.
The hot gas component of
the Galaxy typically has a density of $10^{-2.5}\,
\mbox{cm}^{-3}$ and a temperature of $10^{5.7}$ K (McKee
\& Ostriker 1977).
The hot gas will flow out of the Galaxy (Shapiro \& Field
1976), because its thermal energy is larger than the
gravitational binding
energy of the Galaxy (Cox \& Smith 1974). Using the
calculation by Habe \& Ikeuchi (1980), we estimate the mass
ejection (escaping) rate as
$\dot{M}\sim 1\, M_\odot\,\mbox{yr}^{-1}$.
Assuming a steady, spherically symmetric flow, we can write
approximately
\begin{equation}
\dot{M}\simeq 4\pi R^2\rho v_{\rm esc} ,
\end{equation}
where $R$ is the galactocentric distance, and $\rho$
and $v_{\rm esc}$ are the density and the velocity of the
escaping wind at
distance $R$, respectively (see also Wang 1995).
From equation (1), $\rho$ is expressed as follows:
\begin{equation}
\rho\simeq\frac{\dot{M}}{4\pi R^2v_{\rm esc}}.
\end{equation}

Next we consider a dwarf  galaxy which is located at
the distance $R$ from the Galaxy. Ram pressure by the
wind is $\sim\rho v_{\rm esc}^2$. For the
ram pressure to remove the gas from the dwarf, we require
\begin{equation}
\rho v_{\rm esc}^2\pi r_{\rm s}^2\ga
{\cal E}_{\rm b}/r_{\rm s},
\end{equation}
where $r_{\rm s}$ and ${\cal E}_{\rm b}$ are the surface
radius and the
binding energy of the dwarfs, respectively. $r_{\rm s}$ and
${\cal E}_{\rm b}$ are calculated from observations of
velocity dispersions (Saito 1979a; see also Table 1).
Approximately the left-hand side of (3) represents
the total ram
pressure force on the dwarf, while the right-hand
side represents the total gravitational force. Combining
(2) and (3), we find the following
condition,
\begin{equation}
R^2\la
\frac{\dot{M}r_{\rm s}^3v_{\rm esc}}{4{\cal E}_{\rm b}}
\equiv R_{\rm c}^2.
\end{equation}
Here, $R_{\rm c}$ is the critical radius.
If $R<R_{\rm c}$, ram pressure exceeds binding force so
that the wind can strip the gas from the dwarf.
The velocity of the wind escaping from the Galaxy is
$v_{\rm esc}\simeq 300\,\mbox{km}\,\mbox{s}^{-1}$
(Habe \& Ikeuchi 1980). We finally derive
\begin{eqnarray}
R_{\rm c} &\simeq & 120\left(
\frac{\dot{M}}{1\, M_\odot\,\mbox{yr}^{-1}}
\right)^{1/2}\left(\frac{r_{\rm s}}{1\,\mbox{kpc}}
\right)^{3/2}
\left(\frac{v_{\rm esc}}{300\,\mbox{km}\,\mbox{s}^{-1}}
\right)^{1/2}
\left(\frac{{\cal E}_{\rm b}}{10^{50}\,\mbox{erg}}
\right)^{-1/2}\mbox{kpc}.
\end{eqnarray}

The calculated critical radii are listed in Table
1. Moreover, we calculate and list the crossing time,
$t_{\rm cross}\equiv R/v_{\rm esc}$, which is typically
\begin{equation}
t_{\rm cross}\simeq 0.3
\left(\frac{R}{100\,\mbox{kpc}}\right)
\left(\frac{v_{\rm esc}}{300\,\mbox{km\,s}^{-1}}\right)^{-1}
\,\mbox{Gyr}.
\end{equation}

\section{DISCUSSIONS}

\subsection{Star formation histories}

Observationally, Ursa Minor, Draco and Sextans are
classified as A in Table 1, in which the stellar
population is dominated by early bursts of
star formation. For Ursa Minor and Draco,
we find $R<R_{\rm c}$. Thus it
seems highly probable that the galactic
wind has stripped the gas from these dwarfs within
$\sim 1$ Gyr
after the wind formation in the Galaxy.

The other dwarf galaxies in Table 1 belong to the
category B.
Indeed they satisfy the condition $R>R_{\rm c}$,
and the ram pressure by the wind has little influence on
them. In two most distant dwarfs, particularly, Phoenix and
DDO 210, star formation
is still going on (van den Bergh 1994).
This is because they practically feel no
ram pressure by the wind.

We also comment that
as for dwarf galaxies in the category A, a radial `pumping'
mode, in which mass flows quasi-periodically into the core
of the galaxies (Balsara, Livio, \& O'Dea 1994), is not
permitted, because of their shallow
gravitational potential, although this mode may exist for
more distant dwarfs in the category B (Comins 1984).

\subsection{The Large Magellanic Cloud}

Though the Galactic wind should have a considerable influence on
the Large Magellanic Cloud (LMC), the nearest galaxies from the
Galaxy, there are
some evidences of recent star formation in the LMC
(e.g., Massey 1990). 
Someone might think that the LMC breaks
our model. But, throughout our discussions, we have only discussed
the steady galactic wind. In the realistic situations, however,
supernova rate
(star formation rate) decreases as time goes on. According to Larson
\& Tinsley (1978), its decaying timescale may be about $\sim$ 1 Gyr.
In recent epochs, OB-star formation rate also decreases and then the
Galactic wind becomes weaker.
In other words, the Galactic halo is now bound and cooled type
(Habe \& Ikeuchi 1980), because OB-star heating in the Galactic disk
becomes insignificant [galactic fountain picture like
Shapiro \& Field (1976)]. This makes possible the recent star
formation in the LMC, since
the LMC is
free from the Galactic wind.
Indeed, the LMC has a bimodal age distribution of stellar populations,
with almost all
of the
star clusters in the LMC being either younger than 3 Gyr or older
than 12
Gyr (Da Costa 1991). First star formation is stopped by the Galactic
wind, and second star formation becomes possible after the Galactic
wind has stopped. We should note that the LMC is much larger
gravitational potential than galaxies listed in Table1, which makes
the second star formation easier.

Thus, through our galactic wind model,
we can understand proximity effect on star formation rate in nearby
galaxies.

\subsection{The importance of Galactic wind}

In the previous two subsections, we have verified the
importance of galactic wind in the following three points:

\noindent
[1] Gas in dwarf galaxies located within $R_{\rm c}$
from the Galaxy ($R_{\rm c}$ is defined in Eq. 5) can
be stripped by ram pressure by the
Galactic wind.

\noindent
[2] The process of ram pressure stripping by the Galactic wind
must be considered
in studying star formation histories of Local Group dwarf
galaxies.

\noindent
[3] The star formation histories of the LMC can be explained
consistently by adopting a Galactic wind model (see \S 3.2). 
Thus, our
Galactic wind model is successful in understanding the star
formation histories of Local Group galaxies.

\section{CONCLUSION}

On the basis of the observational results in van den
Bergh (1994), we suggest that star formation histories
of Local Group dwarf galaxies depend on the distance
from the Galaxy. If the distance from the Galaxy is less
than the critical
radius $R_{\rm c}$, which is given in
equation (5), the gas of the dwarf galaxy
is stripped by ram
pressure exerted by the Galactic wind,
since the ram pressure
force by the wind exceeds the gravitational
binding force by the dwarf itself.

Thus we can conclude that galactic wind from
a giant galaxy generally has considerable influence on star
formation histories of nearby dwarf galaxies.

\section*{Acknowledgments}

We thank T. T. Takeuchi for useful discussions and comments.

\end{document}